\documentclass[10pt]{article}

\usepackage{amsmath}
\usepackage{amssymb}

\usepackage{graphicx}
\usepackage{fixltx2e}
\usepackage{hyperref}

\graphicspath{ {figs/} }

\usepackage{cite}

\usepackage{color} 

\usepackage[labelfont=bf,labelsep=period,justification=raggedright]{caption}

\makeatletter
\renewcommand{\@biblabel}[1]{\quad#1.}
\makeatother

\date{}

\begin{document}

\begin{flushleft}
{\Large\textbf{Modeling the Heart as a Communication System}}

\end{flushleft}
\begin{flushleft}
Hiroshi Ashikaga$^{1\ast}$
Jos\'{e} Aguilar-Rodr\'{i}guez$^{2,3}$,
Shai Gorsky$^{4}$,
Elizabeth Lusczek$^{5}$,
Fl\'{a}via Maria Darcie Marquitti$^{6}$,
Brian Thompson$^{7}$,
Degang Wu$^{8}$,
Joshua Garland$^{9}$
\end{flushleft}
\begin{flushleft}
\bf{1} Division of Cardiology, Johns Hopkins University School of Medicine, Baltimore, MD, USA
\\
\bf{2} Institute of Evolutionary Biology and Environmental Studies, University of Zurich, Switzerland
\\
\bf{3} Swiss Institute of Bioinformatics, Lausanne, Switzerland
\\
\bf{4} Department of Economics, University of Utah, Salt Lake City, UT, USA
\\
\bf{5} Department of Surgery, University of Minnesota, Minneapolis, MN, USA
\\
\bf{6} Departmento de Ecologia, Universidade de S\~{a}o Paulo, Brazil
\\
\bf{7} Network Science Division, US Army Research Laboratory, Adelphi, MD, USA
\\
\bf{8} Department of Physics, The Hong Kong University of Science and Technology, Clear Water Bay, Hong Kong, HKSAR, China
\\
\bf{9} Department of Computer Science, University of Colorado, Boulder, CO, USA 
\\
\end{flushleft}
\begin{flushleft}
*E-mail: hashika1@jhmi.edu
\end{flushleft}

\section*{Summary}
Electrical communication between cardiomyocytes can be perturbed during arrhythmia, but these perturbations are not captured by conventional electrocardiographic metrics. We developed a theoretical framework to quantify electrical communication using information theory metrics in 2-dimensional cell lattice models of cardiac excitation propagation. The time series generated by each cell was coarse-grained to 1 when excited or 0 when resting. The Shannon entropy for each cell was calculated from the time series during four clinically important heart rhythms: normal heartbeat, anatomical reentry, spiral reentry, and multiple reentry. We also used mutual information to perform spatial profiling of communication during these cardiac arrhythmias. We found that information sharing between cells was spatially heterogeneous. In addition, cardiac arrhythmia significantly impacted information sharing within the heart. Entropy localized the path of the drifting core of spiral reentry, which could be an optimal target of therapeutic ablation. We conclude that information theory metrics can quantitatively assess electrical communication among cardiomyocytes. The traditional concept of the heart as a functional syncytium sharing electrical information cannot predict altered entropy and information sharing during complex arrhythmia. Information theory metrics may find clinical application in the identification of rhythm-specific treatments which are currently unmet by traditional electrocardiographic techniques.  
  
\section*{Key Index Words }
Information Theory; Cardiac Electrophysiology; Cardiac Arrhythmia; Mathematical Modeling

\newpage
\section*{Introduction}
The human heart consists of 5 billion autonomous cardiomyocytes\cite{Kapoor:2013aa} with simple rules of operation and minimal central control. The behaviors of individual cardiomyocytes are orchestrated by electrical conduction between adjacent cells connected by specialized cell-to-cell junctions called intercalated discs\cite{guyton:2010aa}. The intercalated discs contain gap junctions with large nonselective connexin channels that allow ions and other small molecules to diffuse freely in the cytosol of adjacent cells and reduce internal electrical resistance\cite{Fishman:1990aa}. By providing low-resistance connections between cardiomyocytes, gap junction channels allow electrical waves to propagate rapidly throughout the heart\cite{katz:2010aa}. However, this task can be perturbed during cardiac arrhythmia (abnormal heart rhythm). Normally functioning cardiomyocytes can not only interrupt the electrical information flow at a wavebreak\cite{Pertsov:1993aa}, they can also generate a completely different output by creating a reentry circuit where the wave rotates around the wavebreak. Arrhythmia can result from a wavebreak in an intersection between a wavefront and a wavetail\cite{Weiss:2005aa}, which leads to loss of information about the input.

Conventional electrocardiographic metrics can measure the sequence of electrical excitation\cite{Ashikaga:2007zb, Stevenson:2008ne, Wang:2011ap}, but cannot quantify how arrhythmia impacts the communication between individual cardiomyocytes. In contrast, information theory metrics such as mutual information can quantify the sharing of information in the presence of arrhythmia. Information theory has been used to evaluate biological communication in computational neuroscience\cite{Ruyter-van-Steveninck:1997aa}, transcriptional regulation\cite{Ziv:2007aa,Tkacik:2008aa}, bacterial quorum sensing\cite{Mehta:2009aa}, chemotaxis\cite{Fuller:2010aa}, biochemical signaling networks\cite{Cheong:2011aa}, and evolutionary biology\cite{Adami:2012aa}. We propose a novel application of information theory to analyze the cardiac electrical communication system (Figure \ref{fig:fig_channel_grid.png}A).

We propose that electrical wave propagation is the mechanism by which information is shared between cardiomyocytes in the whole heart. Under this paradigm, heart rhythm disorders result from abnormal production and transmission of information that can be quantified by information theory measures. To this end, we developed a framework to quantify cardiac electrical communication during action potential propagation in normal and abnormal heart rhythms. Three major mechanisms of clinically important arrhythmias that could lead to sudden death were considered: anatomical reentry (reentry around an anatomically defined circuit\cite{Stevenson:1993gy}), spiral reentry (functional reentry without an anatomically defined circuit\cite{Gray:1998aa}), and multiple reentry (multiple functional reentry circuits in the presence of ongoing wavebreak\cite{MOE:1964aa}). 

\section*{Materials and Methods}

\subsection*{Models of Action Potential Propagation}
To develop the information theory framework, we employed two commonly used mathematical models of cardiac action potential propagation. The first model was a monodomain reaction diffusion (RD) model that was originally derived by FitzHugh\cite{Fitzhugh:1961aa} and Nagumo\cite{Nagumo:1962aa} as a simplification of the biophysically based Hodgkin-Huxley equations describing current carrying properties of nerve membranes\cite{Hodgkin:1952aa} which was later modified by Rogers and McCulloch to represent cardiac action potential\cite{Rogers:1994aa}. This model reproduces several physiological properties known to be important in arrhythmogenesis including slowed conduction velocity ($CV$) and unidirectional block due to wavefront curvature\cite{Rogers:1994aa}. This model was used widely in previous studies\cite{Plank:2008aa,Campbell:2009aa,Bourgeois:2009aa,Plank:2009ci,Chamakuri:2013aa,Le:2013aa,Sovilj:2013aa}. 

\begin{eqnarray}
\frac{\partial v}{\partial t}&=&0.26v(v-0.13)(1-v)-0.1vr+ I_{ex}+G_{x}\frac{\partial^2 v}{\partial x^2}+G_{y}\frac{\partial^2 v}{\partial y^2} 
\label{eq:FHN01}
\\
\frac{\partial r}{\partial t}&=& 0.013(v-r)
\label{eq:FHN02}
\end{eqnarray}
Here, $v$ is the excitation variable which can be identified with transmembrane potential, $r$ is the recovery variable, $I_{ex}$ is the external current\cite{Pertsov:1993aa}, and $G_{x}$ and $G_{y}$ are the conductance in $x$ and $y$ directions on the lattice, respectively. In this study the lattice was assumed to be isotropic (i.e., $G_{x}=G_{y}$). The model equations were solved using a finite difference method for spatial derivatives and explicit Euler integration for time derivatives assuming Neumann boundary conditions.

Cellular automata (CA) models have been used to study cardiac action potential propagation in several previous studies\cite{Alonso-Atienza2007aa,Barquero-Perez:2009aa,Beltran-Molina:2011aa,Beltran-Molina:2012aa,Requena-Carrion:2007aa,Requena-Carrion:2013ab,Requena-Carrion:2009aa,Vaisanen:2008aa,Vaisanen:2007aa,Vaisanen:2007ab}. The model we used employed realistic restitution properties and the curvature phenomenon\cite{Alonso-Atienza:2005aa}. Each cell can adopt one of the following three physiologically meaningful states: resting, refractory$_1$ and refractory$_2$. Cells in the resting state are relaxed and can be excited while cells in both refractory states are excited. Cells in refractory$_1$ can excite neighboring cells, while cells in refractory$_2$ cannot. The depolarization (or excitation) of a cell is the transition from the resting state into the refractory$_1$ state and occurs according to a probabilistic update rule $P^{exc}$ based on two influences: (1) the intrinsic cell excitability ($E$) that increases with the time of a cell at rest, and (2) the amount of excitation in the neighborhood of a cell ($Q$):

\begin{equation}
P_{j}^{exc}=EQ=E\sum_{i\neq j}\frac{A_{i}}{d_{ij}^{2}}
\label{eq:CA01}
\end{equation}
where $i$ is a cell adjacent to $j$, $A_{i}$ is a binary excitation state with a value of 0 for the resting state and a value of 1 for either of the two refractory states, and $d_{ij}$ is the distance between the midpoints of cells $i$ and $j$. $E$ was estimated using the restitution curve of the $CV$, which depends on the duration of the previous diastolic interval ($DI$). The transitions from refractory$_1$ to refractory$_2$ (partial repolarization) and from refractory$_2$ to resting (total repolarization) are deterministic. The total time spent in the two refractory states matches the total action potential duration ($APD$). The period in the refractory$_1$ state is equal to 10\% of the $APD$. The $APD$ was estimated based on the restitution curve of the $APD$, which is also a function of the duration of the preceding $DI$. In addition to $APD$ and $CV$ restitution properties, Equation \ref{eq:CA01} also reproduces $CV$ slowing in areas with a pronounced wavefront curvature because of the decreased probability of excitation. The action potential was reproduced according to the Luo-Rudy model\cite{Luo:1991aa}, which associates the time for which each cell has been in its current state with its voltage level.

For both models, the cardiac tissue was simulated as a 2-dimensional (2-D) $128 \times 128$ isotropic lattice of cells (Figure  \ref{fig:fig_channel_grid.png}B). In each cell, the time series of cardiac excitation was computed for 10 seconds with a discrete sampling rate of 500/sec (temporal resolution $\Delta\textit{t}$=2 ms, Figure \ref{fig:fig_channel_grid.png}C). The duration and the sampling rate of the time series were determined to reflect realistic measurements in human clinical electrophysiology studies\cite{fogoros:2012aa}. 

\subsection*{Cardiac Simulation}
We simulated four different heart rhythms in both the RD and CA models: normal heartbeat, anatomical reentry, spiral reentry, and multiple reentry. The latter two are considered to be important mechanisms of cardiac fibrillation, including atrial fibrillation (AF) and ventricular fibrillation (VF)\cite{Gray:1998aa,MOE:1964aa}. Cardiac simulation was performed using MATLAB R2014a (Mathworks, Inc.).

Normal heartbeat was simulated as regular point stimulations (60 beats/min) originating from the top middle region of the lattice. This pattern of stimulation caused a regular train of curved excitation wavefronts traveling from top to bottom along the vertical axis in both the RD (Video S1) and CA models (Video S2).

Anatomical reentry is characterized by an electrical wavefront that travels along a preformed anatomical obstacle, most commonly a scar resulting from healed myocardial infarction, and re-excites previously excited tissue. We simulated a cardiac impulse which can rotate around the obstacle, leading to repetitive, rapid excitation of the heart. Anatomical reentry was reproduced in the RD (Video S3) and the CA models (Video S4) by simulating a non-excitable circular region in the center of the lattice occupying ~20\% of the total surface area.

For spiral reentry\cite{Gray:1998aa}, we simulated a 2-D wave of excitation emitted by an organizing source (or `rotor') of functional reentry whose front is an involute spiral with increasing convex curvature toward the rotation center\cite{:2009sl_ch07}. The spiral reentry was generated by a cross-field stimulation protocol\cite{Pertsov:1993aa} in both the RD (Video S5) and the CA models (Video S6).

Multiple reentry is characterized by multiple independent circuits of functional reentry occurring simultaneously and propagating randomly throughout the cardiac tissue\cite{Calkins:2012aa}. Wavefronts continuously undergo wavefront-wavetail interactions resulting in wavebreak and generation of new wavefronts\cite{Schotten:2011aa}. Multiple reentry was reproduced in the RD (Video S7) and the CA models (Video S8) by a train of random point stimulations in the substrate where the APD is shortened by 40\%. 

\subsection*{Information Measures}

\begin{figure}[!ht]
	\begin{center}
	\includegraphics[width=\textwidth]{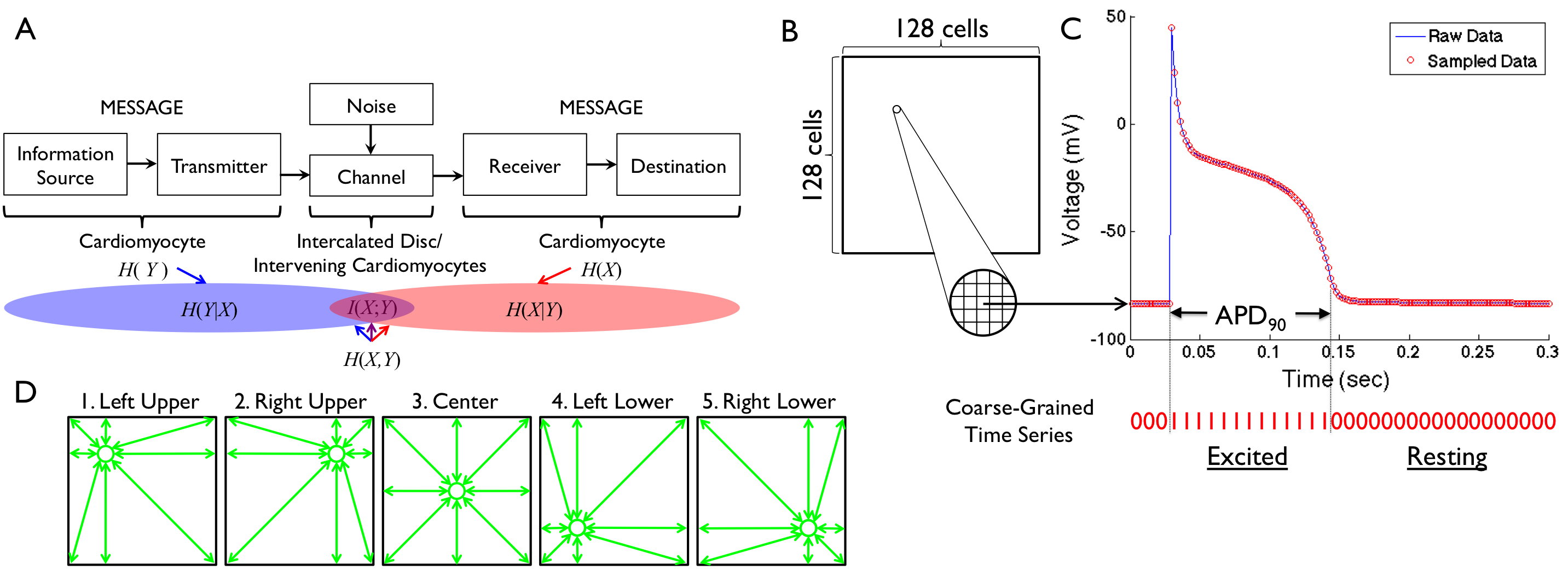}
	\end{center}
	\caption{
	{\bf Conceptual Overview of the Methods.} {\it A. The Heart as a Communication System.} The heart can be considered as a communication system where cardiomyocytes act as an information source/transmitter and a receiver/destination with channels being intercalated discs/intervening cardiomyocytes. $H(X)$ and $H(Y)$ are the entropies of time series $X$ and $Y$ respectively; $H(X,Y)$ is the joint entropy of $X$ and $Y$; $H(X|Y)$ is the conditional entropy of $X$ given $Y$; $H(Y|X)$ is conditional entropy of $Y$ given $X$; $I(X;Y)$ is the mutual information of $X$ and $Y$. Figure modifed from\cite{1948:shannon01}.{\it B. 2-D Lattice Model of Cardiac Tissue}. For both the reaction diffusion and the cellular automata models, the cardiac tissue was simulated as a 128 x 128 cell lattice, which was assumed to be isotropic. {\it C. Coarse-Graining of the Time Series}. In each cell, the time series of cardiac excitation was computed for 10 seconds during four different heart rhythms (normal heartbeat, spiral reentry, anatomical reentry, and multiple reentry) at a sampling rate of 500/sec. The time series was coarse-grained to 1 when excited (during action potential duration at 90\% repolarization \texorpdfstring{APD\textsubscript{90}},) or 0 when resting. {\it D. Mutual Information}. To understand the spatial characteristics of information sharing among cardiomyocytes, mutual information was computed between five representative cells (green circles) and all the other cells in the 2-D lattice. These representative cells included cells in the left-upper quadrant \texttt{(32,32)}, the right-upper quadrant \texttt{(32,96)}, the center \texttt{(64,64)}, the left-lower quadrant \texttt{(96,32)} and the right-lower quadrant \texttt{(96,96)}.}
	\label{fig:fig_channel_grid.png}
\end{figure}

For each cell, the time series of cardiac excitation was coarse-grained to one when excited (during the APD at 90\% repolarization, or APD$_{90}$) or zero when resting (Figure \ref{fig:fig_channel_grid.png}C). We treated each cell on the lattice as a time-series process $X$ where at any observation time $t$ the process $X$ is either excited or resting, in which case we define $X_t = 1$ or $X_t = 0$, respectively.

Using this framework, we can compute the Shannon entropy $H$ of each time-series process $X$:
\begin{equation}
H(X)=-\sum_{x}p(x)\log_{2}p(x)
\label{eq:entropy}
\end{equation}
where $p(x)$ denotes the probability density function of the time series generated by $X$. This quantifies the average uncertainty of whether a single cell is excited or resting over each cell's time history \cite{2006:CoverEIT}.

Mutual information $I(X;Y)$ is a measure of the reduction in uncertainty of the time-series process $X$ due to the information gained from knowing the time-series process $Y$; hence, this quantity is commonly viewed as the information shared between $X$ and $Y$\cite{2006:CoverEIT}. Therefore, by computing the mutual information, we can receive insight into which cells share information and how much. Formally, the mutual information between the time-series processes, in this case cells, $X$ and $Y$ is:

\begin{eqnarray}
I(X;Y)&=&\sum_{x,y}p(x,y)\log_{2}\frac{p(x,y)}{p(x)p(y)}
\label{eq:mi01}
 \\
&=&H(X)+H(Y)-H(X,Y)
\label{eq:mi02}
\end{eqnarray}
where $p(x,y)$ and $H(X,Y)$ denote the joint probability density function and the joint entropy of $X$ and $Y$, respectively (Figure \ref{fig:fig_channel_grid.png}A). 

To understand the spatial profiles of information sharing between cardiomyocytes, mutual information was computed between each of five representative cells (green circles in (Figure \ref{fig:fig_channel_grid.png}D) and all other cells in the 2-D lattice taken individually. These representative cells were defined to be in the left-upper quadrant \texttt{(32,32)}, the right-upper quadrant \texttt{(32,96)}, the center \texttt{(64,64)}, the left-lower quadrant \texttt{(96,32)}, and the right-lower quadrant \texttt{(96,96)}. These points were chosen to avoid artifacts generated by the boundary conditions and point stimulation in the RD model as discussed in the {\it Results} section. Custom programs in Python were used to compute information measures.

\section*{Results}
\subsection*{Normal Heartbeat}
\begin{figure}[!ht]
	\begin{center}
	\includegraphics[width=\textwidth]{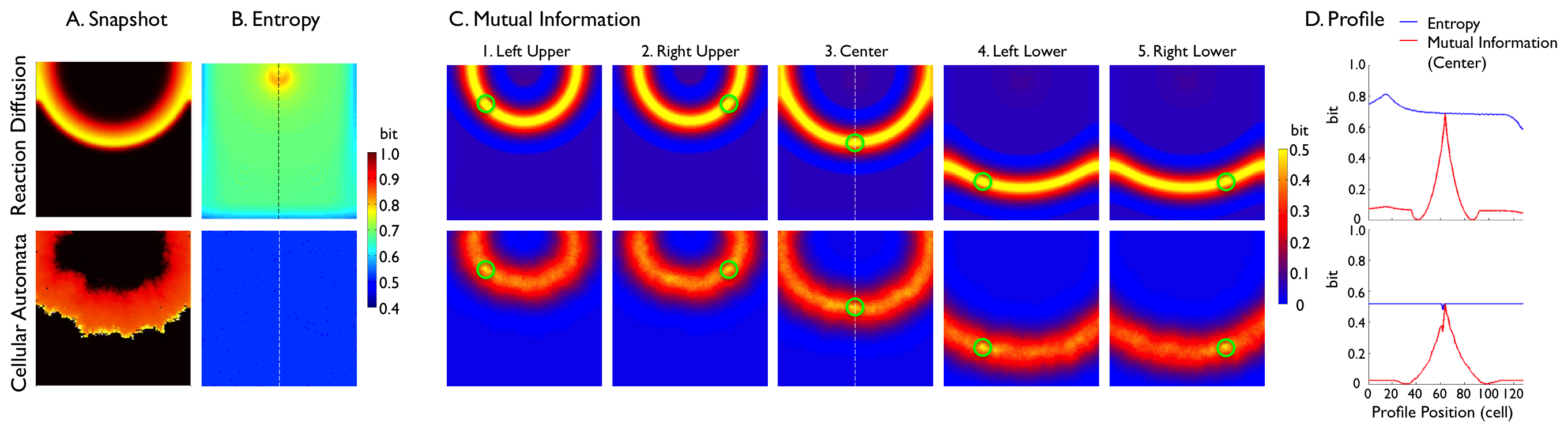}
	\end{center}
	\caption{
	{\bf Normal Heartbeat.} The top row represents the deterministic reaction diffusion (RD) model and the bottom row represents the probabilistic cellular automata (CA) model. {\it A. Snapshot of Electrical Wave Propagation} shows a representative snapshot of an electrical wave from Videos S1 (top) and S2 (bottom). {\it B. Entropy of Each  Cell in the Cardiac Tissue} shows a heat map of the entropy in bits over the cell lattice. {\it C. Mutual Information Between Two Cells in the Cardiac Tissue} shows a heat map of mutual information in bits over the cell lattice. Mutual information was computed between a specific cell (green circle) and all the other cells in the same cardiac tissue. {\it D. Profiles of Entropy and Mutual Information} show profiles of entropy (blue line) and mutual information (red line) through the cell lattice along the vertical broken line shown in {\it 2B} and {\it 2C3. Center}, respectively.}
	\label{fig:fig_ps.png}
\end{figure}

In the RD model, electrical wavefronts regularly swept the lattice from top to bottom (Video S1, Figure \ref{fig:fig_ps.png}A, top row). The entropy was relatively lower at the lattice borders and higher at the site of stimulation, but these were artifacts of the boundary conditions and point stimulation, respectively (Figure \ref{fig:fig_ps.png}B, top row). Otherwise, entropy was homogeneous across the lattice [0.68 (mean)$\pm$ 0.03 (SD) bits]. Mutual information showed a spatially heterogeneous information sharing between cells (Figure \ref{fig:fig_ps.png}C, top row) despite the assumed isotropic structure and homogeneous electrical properties of the lattice. For example, Figure \ref{fig:fig_ps.png}C3 (top row) shows that the cell in the center of the lattice shares a high amount of information with the cells on the same electrical wavefront (yellow band) generated by the heartbeats. It also shows little information sharing with the cells preceding and following the wavefront (light blue bands surrounding the yellow band). This is clearly shown in the profile of mutual information (Figure \ref{fig:fig_ps.png}D, top row) along the vertical broken line in Figure \ref{fig:fig_ps.png}C3 (top row). Mutual information (red line) reached its peak (0.69 bits) at the center (profile position 64), where mutual information is equal to entropy because the mutual information of an entity with itself is equal to entropy (Equation \ref{eq:mi02}). Mutual information fell off sharply from the center and reached the minimum (0 bits) on both sides (profile position 41 and 86) before slightly rising to approximately 0.06 bits on both ends. These findings indicate that the cells share a high amount of information when they are in phase with the cardiac excitation and share little information when they are out of phase. We formed an analytical framework which corroborates these numerical results (Appendix S1).

In the CA model (Video S2, Figure \ref{fig:fig_ps.png}A, bottom row), the electrical wavefront was more irregular and unstable than that of the RD model. Entropy was homogeneous across the lattice, but was lower ($0.52 \pm 0.00$ bits) than that of the RD model (Figure \ref{fig:fig_ps.png}B, bottom row). This difference results from the fact that the CA model had a longer resting state (Video S2), making it more biased towards the resting state than the RD model (Video S1). There was no qualitative difference in information sharing between the RD model (Figure \ref{fig:fig_ps.png}, top row) and the CA model (Figure \ref{fig:fig_ps.png}, bottom row). However, information sharing was lower in the CA model because the CA model is inherently probabilistic and less reproducible than the RD model. The profile of mutual information (Figure\ref{fig:fig_ps.png}D) along the vertical broken line in Figure \ref{fig:fig_ps.png}C3 also shows a qualitatively similar but lower mutual information in the CA model (Figure \ref{fig:fig_ps.png}D) relative to the RD model (Figure \ref{fig:fig_ps.png}D).

\subsection*{Anatomical Reentry}
\begin{figure}[!ht]
	\begin{center}
	\includegraphics[width=\textwidth]{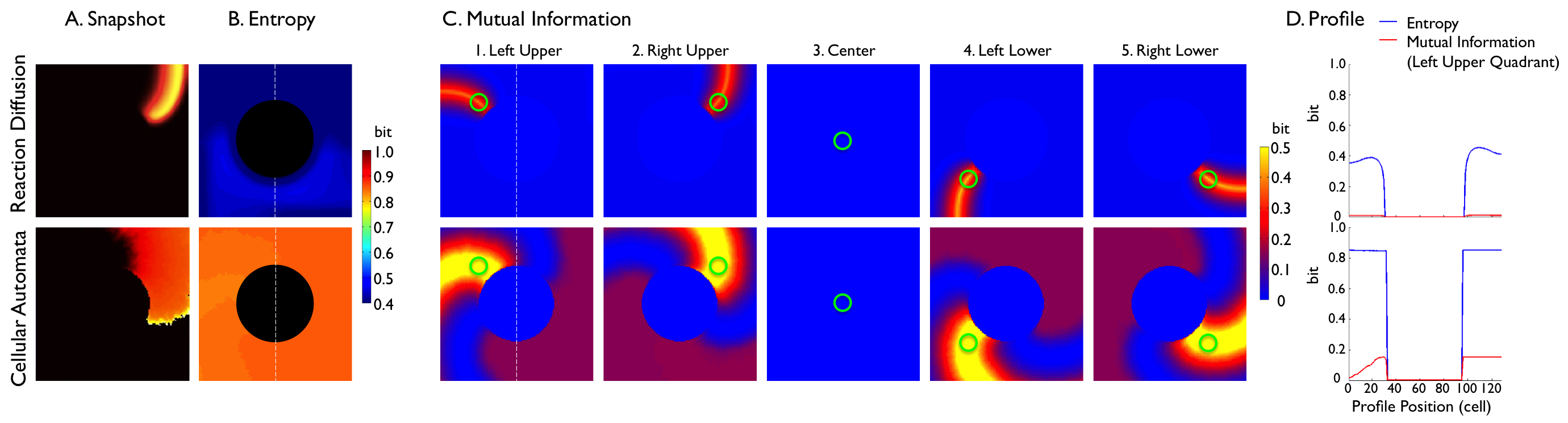}
	\end{center}
	\caption{
	{\bf Anatomical Reentry.} The black circular region in A and B represents a non-excitable tissue that serves as an anatomical obstacle around which cardiac excitation rotates perpetually. The top row represents the deterministic reaction-diffusion (RD) model and the bottom row represents the probabilistic cellular automata (CA) model. {\it A. Snapshot of Electrical Wave Propagation} shows a representative snapshot of an electrical wave from Videos S5 (top) and S6 (bottom). {\it B. Entropy of Each  Cell in the Cardiac Tissue} shows a heat map of the entropy in bits over the cell lattice. {\it C. Mutual Information Between Two Cells in the Cardiac Tissue} shows a heat map of mutual information in bits over the cell lattice. Mutual information was computed between a specific cell (green circle) and all the other cells in the same cardiac tissue. {\it D. Profiles of Entropy and Mutual Information} shows the profile of entropy (blue line) through the cell lattice along the vertical broken line shown in {\it 4B}. The profile of mutual information (red line) in {\it 4D} is from {\it 4C1. Left Upper Quadrant}.}
	\label{fig:fig_ar.png}
\end{figure}

In both the RD (Video S3) and the CA models (Video S4), the entropy of the cells within the circular non-excitable region was zero because these cells were always in the resting state (Figure \ref{fig:fig_ar.png}A, B). The entropy was roughly homogeneous in other regions of the lattice. The average entropy was 0.32 $\pm$ 0.17 bits in the RD model and 0.69 $\pm$ 0.33 bits in the CA model (Figure \ref{fig:fig_ar.png}B). The difference resulted from the longer wavelength in the CA model due to the more convex curvature than the RD model.

Overall, both models showed a similar spatial pattern of mutual information (Figure \ref{fig:fig_ar.png}C). Information sharing between cells was spatially heterogeneous but showed rotational symmetry about the non-excitable region in the center. For example, Figure \ref{fig:fig_ar.png}C1 shows that the cell in the left upper quadrant of the lattice shares a high amount of information with the cells on the same electrical wavefront (the orange band in the RD model and the yellow band in the CA model). Information sharing in the cells on the three other quadrants (Figure \ref{fig:fig_ar.png}C2, 4 and 5) was rotationally symmetric with that of Figure \ref{fig:fig_ar.png}C1. Importantly, there is no information sharing between the cell within the circular non-excitable region and any other cells in the lattice (Figure \ref{fig:fig_ar.png}C3). This is logical from both the standpoints of electrophysiology and information theory. Of note, similar to normal heartbeat (Figure \ref{fig:fig_ps.png}C), both models also showed little information sharing with the cells that preceded and followed the region of a high amount of shared information (light blue bands before and after the yellow band). These findings indicate that, similar to normal heartbeat, the cells share a high amount of information when they are in phase with cardiac excitation, and share little information when they are out of phase. 

\subsection*{Spiral Reentry}
\begin{figure}[!ht]
	\begin{center}
	\includegraphics[width=\textwidth]{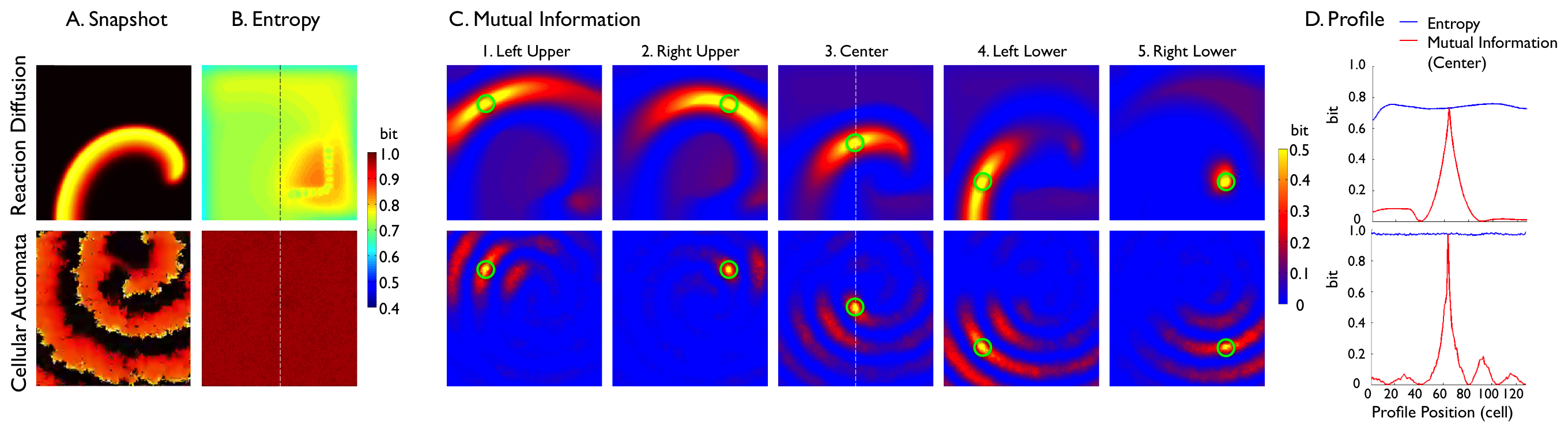}
	\end{center}
	\caption{
	{\bf Spiral Reentry.} The top row represents the deterministic reaction diffusion (RD) model and the bottom row represents the probabilistic cellular automata (CA) model. {\it A. Snapshot of Electrical Wave Propagation} shows a representative snapshot of an electrical wave from Videos S3 (top) and S4 (bottom). {\it B. Entropy of Each  Cell in the Cardiac Tissue} shows a heat map of the entropy in bits over the cell lattice. {\it C. Mutual Information Between Two Cells in the Cardiac Tissue} shows a heat map of mutual information in bits over the cell lattice. Mutual information was computed between a specific cell (green circle) and all the other cells in the same cardiac tissue. {\it D. Profiles of Entropy and Mutual Information} show profiles of entropy (blue line) and mutual information (red line) through the cell lattice along the vertical broken line shown in {\it 3B} and {\it 3C3. Center}, respectively.}
	\label{fig:fig_sw.png}
\end{figure}

In the RD model (Video S5, Figure \ref{fig:fig_sw.png}A, top row), a single rotor with a resultant spiral reentry was simulated in the lattice. The entropy of each individual cell in this simulation shows an important finding with potential clinical significance. The left lower quadrant and both upper quadrants of the lattice exhibited homogeneous entropy ($0.74 \pm 0.04$ bits), except for the borders of the lattice which is an artifact of boundary conditions (Figure \ref{fig:fig_sw.png}B, top row). The red region in the right lower quadrant represents the higher entropy near the spiral tip (rotor) caused by the conduction velocity ($CV$) slowing near the rotor due to a pronounced wavefront curvature (Figure \ref{fig:fig_sw.png}B, top row). This slowing of $CV$ effectively caused longer cardiac excitations, making the cells in this region more biased towards the excited state than the rest of the cells in the lattice not directly affected by the rotor, boosting the entropy in this region. Within the red region is an L-shaped, light green, beadlike structure representing the lower entropy in the path of the drifting core of the spiral reentry around which the rotor revolved. This result suggests that the entropy of individual cells may be used to aid in localizing the core of spiral reentry for therapeutic purposes. Of note, the L-shaped path of the drifting core in this model was artificially determined by the boundary condition of the model.

While the entropy of individual cells provided very important findings, the average entropy over all cells was less informative. In fact, the average entropy between normal heartbeat and spiral reentry in the RD model was very similar (0.68 $\pm$ 0.03 $vs.$ 0.74 $\pm$ 0.04 bits, respectively). This suggests that the spatial profiles of entropy are more useful in highlighting the difference in dynamics than the aggregate information over the entire tissue, as averaging effectively filters out the important features of the entropy landscape.

Mutual information was far more spatially heterogeneous than anticipated from the electrical wave propagation. For example, Figure \ref{fig:fig_sw.png}C3 (top row) shows that the high level of information sharing in the central region of the lattice quickly faded as distance between cells increased. This reflects the fact that the cells lying along the same electrical wavefront changed over time due to the drifting core. This resulted in the smaller region of high information sharing in the spiral reentry than in normal heartbeat. This finding indicates that mutual information can sensitively detect regional heterogeneity of cardiac excitation in spiral reentry, which is not apparent from electrical wave propagation. Of note, information sharing in the right lower quadrant was limited to a focal region without a spiral tail (Figure \ref{fig:fig_sw.png}C5, top row). This is because the cell in the right lower quadrant (green circle) happened to lie on the path of the drifting core, which coincided with a void of cardiac excitation.

In the CA model (Video S6, Figure \ref{fig:fig_sw.png}A, bottom row), the electrical wavefront was inherently more irregular than that of the RD model. The entropy was roughly homogeneous across the lattice and close to 1 (0.97 $\pm$ 0.00 bits) (Figure \ref{fig:fig_sw.png}B, bottom row). The higher entropy of the CA model compared with the RD model was, as in the anatomical reentry, caused by the longer wavelength in the CA model due to the more convex curvature than the RD model (Figure \ref{fig:fig_sw.png}A, bottom row). Unlike the RD model, the entropy in the CA model did not show the path of the drifting core, indicating that the drift trajectory was much more random compared to that of the RD model with respect to the time frame (10 sec) of data acquisition. Information sharing between cells was spatially heterogeneous (Figure \ref{fig:fig_sw.png}C, D) because of the regional heterogeneity of cardiac excitation in spiral reentry. Overall, the CA model showed lower information sharing than the RD model due to the probabilistic nature of the model. 

\subsection*{Multiple Reentry}
\begin{figure}[!ht]
	\begin{center}
	\includegraphics[width=\textwidth]{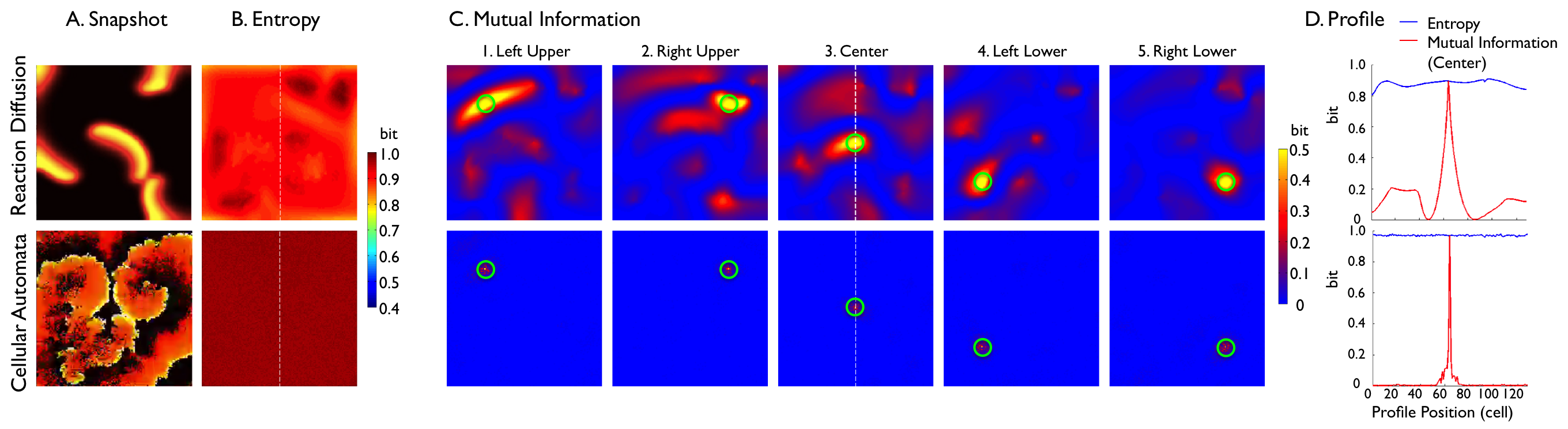}
	\end{center}
	\caption{
	{\bf Multiple Reentry.} The top row represents the deterministic reaction diffusion (RD) model and the bottom row represents the probabilistic cellular automata (CA) model. {\it A. Snapshot of Electrical Wave Propagation} shows a representative snapshot of an electrical wave from Videos S7 (top) and S8 (bottom). {\it B. Entropy of Each  Cell in the Cardiac Tissue} shows a heat map of the entropy in bits over the cell lattice. {\it C. Mutual Information Between Two Cells in the Cardiac Tissue} shows a heat map of mutual information in bits over the cell lattice. Mutual information was computed between a specific cell (green circle) and all the other cells in the same cardiac tissue. {\it D. Profiles of Entropy and Mutual Information} show profiles of entropy (blue line) and mutual information (red line) through the cell lattice along the vertical broken line shown in {\it 5B} and {\it 5C3. Center}, respectively.}
	\label{fig:fig_mw.png}
\end{figure}

In both the RD (Video S7) and the CA models (Video S8), the entropy was homogeneous across the cell lattice (Figure \ref{fig:fig_mw.png}A). The average entropy was 0.88 $\pm$ 0.03 bits in the RD model and 0.97 $\pm$ 0.00 bits in the CA model (Figure \ref{fig:fig_mw.png}B). This indicates that excited and resting states are almost equally distributed throughout the time series of all the cells, yielding a high uncertainty and homogeneous entropy. The spatial profiles of information sharing for both the RD and the CA models were similar to that of spiral reentry (Figure \ref{fig:fig_mw.png}C), except the fact that the underlying structure was much less organized due to the random nature of multiple reentry. Information sharing was low except for a small region in the immediate neighborhood of the cell in which mutual information was measured. Outside this small region information sharing steeply fell off to near zero (Figure \ref{fig:fig_mw.png}D). The near-zero mutual information indicates that the cells almost completely lost synchrony during multiple reentry; that is, individual cardiomyocytes got excited independently from each other and did not share information with cells beyond their immediate neighborhood.

\section*{Discussion}
\subsection*{Summary of the Findings}
By treating the heart as an electrical communication system, we demonstrated quantitatively that information sharing between cardiomyocytes on an isotropic lattice structure is spatially heterogeneous. This finding was unexpected from the traditional concept of the heart as a functional syncytium sharing electrical information via gap junctions, where one might mistakenly assume that information sharing would be homogeneous along the electrical wavefront. We also found that entropy can be significantly different between heart rhythms with electrically similar spatial patterns (Figure \ref{fig:fig_ps.png}B,\ref{fig:fig_sw.png}B and\ref{fig:fig_ar.png}B). These findings indicate that metrics from information theory can quantitatively assess the communication processes within the heart which are not obvious from conventional electrocardiographic metrics such as sequences of electrical excitation. In addition, our results show that cardiac arrhythmia significantly impacts electrical communication within the heart. 

\subsection*{Mutual Information to Quantify Communication Within the Heart}
Analysis of dynamical multivariate data sets over the dimensions of time and physical space is commonly encountered in the investgation of complex systems\cite{galka:2006aa}. The study of cardiac arrhythmia, particularly cardiac fibrillation, is no exception. For example, a number of measures to quantify spatial complexity of VF have been proposed, including the correlation length\cite{Bayly:1993aa}, the multiplicity index\cite{Rogers:1997aa}, and Karhunen-Lo\`eve decomposition\cite{Bayly:1998aa}. The main focus of interest in these studies was to quantify the determinism and the predictability of the time series over physical space.

Our study is different from these previous studies in two aspects. First, our focus of interest was to quantify communication within the heart. Of central importance to the understanding of complex systems is connectivity, or the presence of dynamical interactions between spatially distinct locations within the system. Knowledge about connectivity in a system, whether anatomical or functional, further facilitates the fundamental understanding of the system since it addresses an important aspect of the functional interdependency of between each component of the system. Our results indicate that information theory metrics can quantitatively assess electrical communication processes among cardiomyocytes during normal heartbeat and complex arrhythmias beyond electrocardiographic measures, conferring validity to the paradigm of the heart as a communication system. Second, we used mutual information to perform spatial profiling of different cardiac arrhythmias. Correlations within multivariate time series can be described by measures such as linear and nonlinear correlation functions. However, mutual information has attracted considerable attention recently since it promises a very general quantification of statistical dependence\cite{Li:1990aa}. In addition, previous studies measured the spatial profile of mutual information during VF\cite{Choi:2003aa}, which could include both spiral reentry and multiple reentry, because it is challenging to distinguish one from the other in experimental settings. Therefore, the spatial profile of spiral reentry and multiple reentry was not clearly delineated. Our result showed that the spatial profile and the underlying structure of mutual information during each arrhythmia are clearly different (Figure \ref{fig:fig_sw.png} and \ref{fig:fig_mw.png}). This suggests that information theory metrics may be able to help distinguish one rhythm from another by quantifying communication within the heart.

The underlying mechanism of perturbation of information transfer during arrhythmia remains unclear. Information sharing during VF seems to be directly affected by the anisotropy of myofiber orientation and cell-to-cell coupling\cite{Choi:2003aa}. However, the spatial profile of membrane potential during VF has no consistent relationship with that of intracellular calcium dynamics\cite{Omichi:2004aa}. This may suggest a contribution of non-voltage-gated intracellular calcium release in perturbation of information transfer by increasing the local complex interactions between calcium dynamics and membrane potential. Clearly, further studies will be needed to investigate the mechanistic basis of the paradigm of the heart as a communication system.

\subsection*{Clinical Implications}
Recently, targeted elimination of the rotor (phase singularity) of spiral reentry has been shown to result in sustained termination of AF\cite{Narayan:2012bi}. As a result, spatial localization of the rotor has attracted substantial attention in clinical cardiac electrophysiology. Entropy has been quantified to identify the location of the rotor of spiral reentry from the bipolar electrograms by creating the probability density function based on the amplitude of the signal\cite{Ganesan:2013aa}. However, the accuracy of this metric was not clear, because it has consistent correlation with complex fractionated electrograms\cite{Ng:2010aa}, which was found to bear no spatial relevance to spiral reentry\cite{Narayan:2013aa}. Therefore, the knowledge of the spatial profile of entropy for spiral reentry was lacking. 

Our result in the RD model clearly showed that the region of rotor drift has high entropy (red region, right lower quadrant in Figure \ref{fig:fig_sw.png}B, top row), which is consistent with the previous studies\cite{Ganesan:2013aa}. Moreover, what was most striking was the fact that entropy can localize the path of the drifting core of spiral reentry (L-shaped, light green, beadlike structure, right lower quadrant in Figure \ref{fig:fig_sw.png}B, top row), because of the low entropy of the spiral core. This makes electrophysiological sense because the cardiomyocytes within the spiral core are almost constantly depolarized\cite{Qu:2000mb}, making the probability density biased towards one. Therefore, our result showed a critically important fact that entropy can spatially localize the core (low entropy) within a larger region of the drifting rotor (high entropy). However, because a similar structure of the spatial profile could not be identified in the CA model (Figure \ref{fig:fig_sw.png}B, bottom row), localization of the drifting core may require a spatially stable spiral reentry with adequately slow drift. Although these preliminary findings need to be confirmed in experimental models of cardiac fibrillation, they illustrate the potential clinical utility of information theory applied to cardiac electrophysiology.

\subsection*{Limitations}
There are two limitations that should be considered before our results can be translated to human patients. First, the cardiac tissue was assumed to be a 2-D, isotropic, and homogeneous lattice, whereas real cardiac tissue is 3-D, anisotropic, and heterogeneous due to the intricately woven myofiber structure\cite{Streeter:1969av} and regional heterogeneity\cite{Antzelevitch:2001nk}. These tissue properties may contribute critically to the generation of cardiac arrhythmia\cite{Qu:2000aa}. However, the main focus of this work was to prove the concept that quantitative analysis of electrical communication during existing cardiac arrhythmia could yield clinically relevant results. We used two widely accepted models of action potential propagation in cardiac tissue to reproduce a variety of heart rhythms that captured important features of clinically representative arrhythmias. Therefore, we believe that these model assumptions were acceptable within the scope of this work. Second, our computation of information theory metrics did not incorporate conduction delay of electrical current to travel from one cell to another, since the time series of the entire lattice was acquired simultaneously. This is because the conduction delay within the small 2-D lattice would be negligibly small relative to the acquisition period of 10 seconds. However, this assumption may have underestimated the true amount of information sharing between heart cells because the standard definition of mutual information does not include shared information that is delayed in time. This leaves open the potential for even more clinically useful results by considering generalizations of the metric that explicitly account for this conduction delay.

\subsection*{Conclusions}
Information theory metrics can quantitatively assess electrical communication processes among cardiomyocytes during normal heartbeat and complex arrhythmias beyond electrocardiographic measures. Further, entropy may have a clinical application in the localization and elimination of spiral reentry cores. These results suggest that the heart as a communication system is more complex than the traditional concept of functional syncytium sharing electrical information via gap junctions. We believe that this new paradigm provides a new set of tools for the systems-approach to the heart as a complex system\cite{McCulloch:2005aa}.

\section*{Acknowledgments}
This work was conducted as a research project during the Complex Systems Summer School at the Santa Fe Institute (SFI), NM, USA. The authors thank all the staff at SFI, particularly Sander Bais, Juniper Lovato, and John-Paul Gonzales. The authors also thank Felipe Alonso Atienza, Ferney Beltr\'{a}n-Molina Jes\'{u}s Requena Carri\'{o}n and Peter Hammer for generously providing the source code for the models described in the paper. The authors also thank Simon DeDeo and Nix Barnett for valuable input.

\bibliography{references}
\newpage
\section*{Figure Captions}

\newpage

\newpage

\newpage

\newpage
\section*{Electronic Supplementary Materials}
\begin{flushleft}
{\bf Video S1. Normal Heartbeat (Reaction Diffusion Model).}

{\bf Video S2. Normal Heartbeat (Cellular Automata Model).}

{\bf Video S3. Anatomical Reentry (Reaction Diffusion Model).}

{\bf Video S4. Anatomical Reentry (Cellular Automata Model).}

{\bf Video S5. Spiral Reentry (Reaction Diffusion Model).}

{\bf Video S6. Spiral Reentry (Cellular Automata Model).}

{\bf Video S7. Multiple Reentry (Reaction Diffusion Model).}

{\bf Video S8. Multiple Reentry (Cellular Automata Model).}

{\bf Appendix S1. Sensitivity Analysis of Mutual Information in Cardiac Tissue.}

\end{flushleft}

\section*{Short Title For Page Headings}
The Heart as a Communication System


\end{document}


\begin{center}
{\Large\text{Electronic Supplementary Material for}}
\end{center}

\begin{center}
{\Large\textbf{Modeling the Heart as a Communication System}}
\end{center}

\begin{center}
Hiroshi Ashikaga*, Jos\'{e} Aguilar-Rodr\'{i}guez, Shai Gorsky, Elizabeth Lusczek, Fl\'{a}via Maria Darcie Marquitti, Brian Thompson, Degang Wu, Joshua Garland
\end{center}

\begin{center}
*To whom correspondence should be addressed. E-mail: hashika1@jhmi.edu
\end{center}

\newpage

\begin{center}
{\Large\textbf{Appendix S1. Sensitivity Analysis of Mutual Information in Cardiac Tissue}}
\end{center}

\beginsupplement

In this section, we build an analytical model to determine the sensitivity of mutual information to each parameter of the cardiac tissue using a simple model of unidirectional plane wave propagation. The analytical result should help the reader understand the numerical results in the paper.\\

\subsection*{Basic Cardiac Electrtophysiology}
\begin{figure}[!ht]
	\begin{center}
	\includegraphics[clip,width=5.0cm]{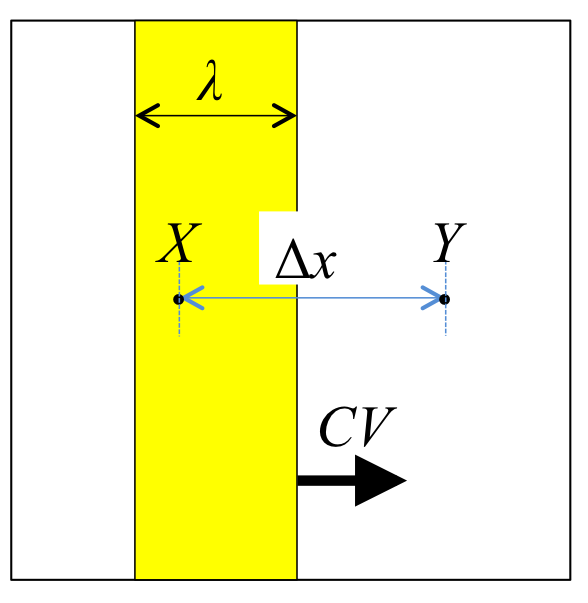}
	\end{center}
	\caption{\footnotesize{{\bf A 2-D Isotropic Excitable Tissue of an Arbitrary Size.} A vertical plane wave of excitationn (in yellow) emerges from the left edge of the tissue, travels toward the right edge with conduction velocity $CV$. $\lambda$ is the wavelength of the excitation wave. $\Delta x$ is the horizontal distance between two arbitrary points $X$ and $Y$ on the cardiac tissue.}}
	\label{fig:yellowband}
\end{figure}

The cardiac tissue is simulated as a 2-dimensional (2-D), isotropic lattice of an arbtrary size (Figure\ref{fig:yellowband}), where a vertical plane wave of excitation emerges from the left edge of the tissue at a constant inter-beat interval (basic cycle length, $BCL$), and travels toward the right edge with conduction velocity $CV$.

\begin{figure}[!ht]
	\begin{center}
	\includegraphics[width=\textwidth]{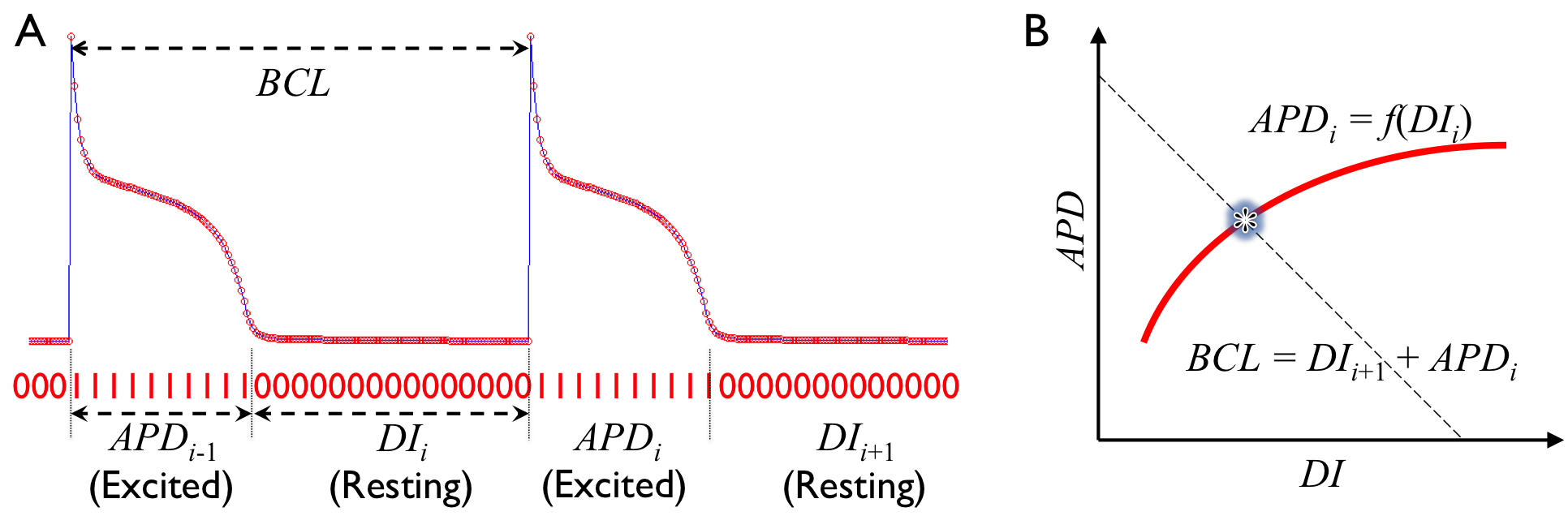}
	\end{center}
	\caption{\footnotesize{{\bf Action Potential Duration (APD) Restitution.} {\it A. Basic cycle length (BCL)}. $BCL$ is the inter-beat interval. $APD$ is the action potential duration and $DI$ is the diastolic interval, where $BCL=DI_{i+1}+APD_{i}$. $i$ denotes the beat number. In our study the time series at each point in the cardiac tissue was coarse-grained to 1 during $APD$ (= excited) or 0 during $DI$ (= resting).  {\it B. APD restitution curve}. $APD$ is a function of the immediately preceding $DI$ ($APD_{i}=f(DI_{i})$). The dashed line represents $BCL=DI_{i+1}+APD_{i}$. The asterisk (*) denotes the stable fixed point when $BCL$ is constant and the slope of the restitution curve at the fixed point is less than one.}}
	\label{fig:bclapddi}
\end{figure}

We define the $BCL$ in terms of the diastolic interval $DI$ and the action potential duration $APD$ (Figure\ref{fig:bclapddi}A):

\begin{eqnarray}
BCL=DI_{i+1}+APD_{i}
\label{eq:bcl}
\end{eqnarray}
where $i$ is the beat number. In this analysis, we assume that $APD$ is determined only by the immediately preceding $DI$ ($APD$ restitution)\cite{Nolasco:1968aa}: 
\begin{eqnarray}
APD_{i}=f(DI_{i})
\label{eq:apdres}
\end{eqnarray}
Other factors could influence $APD$ in more complex models\cite{Berger:2004aa}.

When $BCL$ is constant and the slope of the restitution curve at the fixed point denoted by the asterisk (*) is less than one, that is, $df/dDI<1$, $APD$ and $DI$ converge to constant values (Figure\ref{fig:bclapddi}B). Therefore, the solution to the restitution function at a steady state is 
\begin{eqnarray}
APD=f(DI)
\label{eq:apdresfixed}
\end{eqnarray}
Similar to $APD$, $CV$ in cardiac tissue is a function of the preceding $DI$ and tends to decrease monotonically as $DI$ decreases\cite{2014zipes:aa}.
\begin{eqnarray}
CV_{i}=g(DI_{i})
\label{eq:cvres}
\end{eqnarray}
$CV$ is highest for fully recovered tissue, and there is a minimum DI for propagation at a finite $CV$.  $APD$ and $CV$ are related by the wavelength of the excitation wave $\lambda$:
\begin{eqnarray}
\lambda=APD\times CV
\label{eq:lambda}
\end{eqnarray}

\subsection*{Mutual Information in Cardiac Tissue}
\begin{figure}[!ht]
	\begin{center}
	\includegraphics[clip,width=10.0cm]{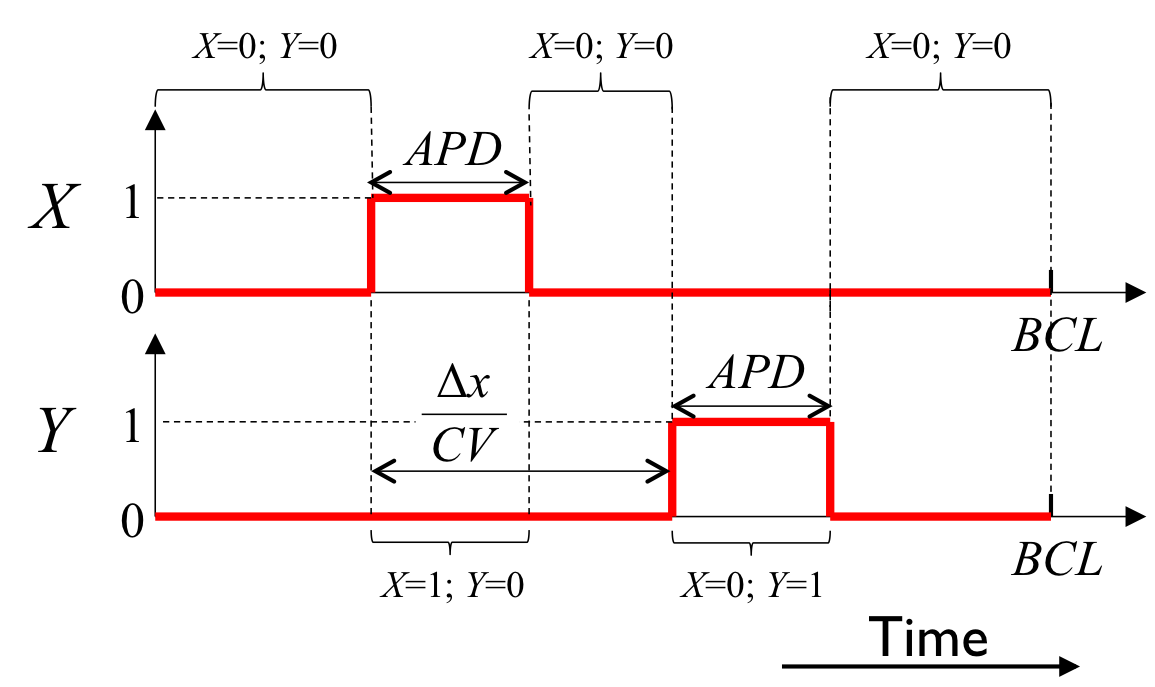}
	\end{center}
	\caption{\footnotesize{\bf Electrophysiological Behaviors of Two Arbitrary Points X and Y in Figure \ref{fig:yellowband}.} This figure illustarates the case where $\Delta x\geq\lambda$, but a similar analysis can be conducted in the case of $\Delta x<\lambda$. The values 0 and 1 represent the 'resting' and 'excited' states, respectively. At $t=0$, the point $X$ is resting. When the plane wave of excitation reaches the point $X$ at the conduction velocity $CV$ it becomes excited, remains excited for the duration of $APD$ (action potential duration), then becomes resting again. The point $Y$, which is located $\Delta{x}$ distal to the point $X$ with respect to the wave propagation, becomes excited $\Delta{x}/CV$ after the point $X$ becomes excited. The duration of excitation at the point Y is also $APD$. This time sequence repeats itself at the rate of $BCL$ (basic cycle length).}
	\label{fig:rect}
\end{figure}

At time $t=0$, the point $X$ in Figure \ref{fig:yellowband} is resting. The point $X$ becomes excited when the plane wave of excitation reaches the point $X$. Therefore, the electrophysiological behaviors of point $X$ over time can be described using a square wave (Figure \ref{fig:rect}). The point $X$ remains excited while the plane wave travels from left to right at the conduction velocity $CV$. From Equation \ref{eq:lambda}, the duration of excitement is
\begin{eqnarray}
\frac{\lambda}{CV}=APD
\label{eq:lambda2}
\end{eqnarray}

When $\Delta x\geq\lambda$, the joint probability distribution $p(x,y)$ at the points $X$ and $Y$ is
\begin{align*}
	&p(X=1,Y=0)=p(X=0,Y=1)=\frac{APD}{BCL}=\frac{APD}{APD+DI}=\frac{1}{1+\frac{DI}{APD}}=\frac{1}{1+\frac{DI}{f(DI)}} 
	\\
	&p(X=1,Y=1)=0
	\\
	&p(X=0,Y=0)=1-\frac{2APD}{BCL}=1-\frac{2}{1+\frac{DI}{f(DI)}} 
\end{align*}

The marginal probabilities $p(x)$ and $p(y)$ at the points $X$ and $Y$ are
\begin{align*}
	&p(X=1)=p(Y=1)=\frac{APD}{BCL}=\frac{1}{1+\frac{DI}{f(DI)}} 
	\\
	&p(X=0)=p(Y=0)=1-\frac{APD}{BCL}=1-\frac{1}{1+\frac{DI}{f(DI)}} 
\end{align*}

Mutual information $I(X;Y)$ is defined as\cite{2006:CoverEIT}
\begin{eqnarray}
I(X;Y)=\sum_{x,y}p(x,y)\log_{2}\frac{p(x,y)}{p(x)p(y)}
\label{eq:mi01}
\end{eqnarray}

If we define a new variable $l$ such that
\begin{eqnarray}
l=\frac{APD}{BCL}=\frac{\lambda}{CV\times BCL}=\frac{1}{1+\frac{DI}{f(DI)}} 
\label{eq:el}
\end{eqnarray}

then the joint probability distribution $p(x,y)$ and the marginal probabilities $p(x)$ and $p(y)$ are
\begin{align*}
	&p(X=1,Y=0)=p(X=0,Y=1)=l
	\\
	&p(X=1,Y=1)=0
	\\
	&p(X=0,Y=0)=1-2l
	\\
	&p(X=1)=p(Y=1)=l
	\\
	&p(X=0)=p(Y=0)=1-l
\end{align*}

Note that $l$ is dependent only on $DI$. From Equation \ref{eq:mi01}, mutual information $I(X;Y)$ is

\begin{align}
I(X;Y) = &p(X=1,Y=0)\log_{2}\frac{p(X=1,Y=0)}{p(X=1)p(Y=0)}+p(X=0,Y=1)\log_{2}\frac{p(X=0,Y=1)}{p(X=0)p(Y=1)} \nonumber \\
 & +p(X=1,Y=1)\log_{2}\frac{p(X=1,Y=1)}{p(X=1)p(Y=1)}+p(X=0,Y=0)\log_{2}\frac{p(X=0,Y=0)}{p(X=0)p(Y=0)} \nonumber \\
= & l\log_{2}\frac{l}{l(1-l)}+l\log_{2}\frac{l}{l(1-l)}+0+(1-2l)\log_{2}\frac{1-2l}{(1-l)^2} \nonumber \\
= & -2(1-l)\log_{2}(1-l)+(1-2l)\log_{2}(1-2l) \label{eq:xlarge}
\end{align}

We define a new variable $d$ such that
\begin{eqnarray}
d=\frac{\frac{\Delta x}{CV}}{BCL}=\frac{\Delta x}{CV\times BCL}=\frac{\Delta x}{g(DI)\left[f(DI)+DI\right]}
\label{eq:dee}
\end{eqnarray}

Note $d$ is dependent only on $DI$ and $\Delta x$. When $\Delta x<\lambda$, the joint probability distribution $p(x,y)$ and the marginal probabilities $p(x)$ and $p(y)$ are

\begin{align*}
	&p(X=1,Y=0)=p(X=0,Y=1)=\frac{\frac{\Delta x}{CV}}{BCL}=d
	\\
	&p(X=1,Y=1)=\frac{APD-\frac{\Delta x}{CV}}{BCL}=l-d
	\\
	&p(X=0,Y=0)=1-\frac{APD+\frac{\Delta x}{CV}}{BCL}=1-l-d
	\\
	&p(X=1)=p(Y=1)=l
	\\
	&p(X=0)=p(Y=0)=1-l
\end{align*}

Therefore, mutual information $I(X;Y)$ is
\begin{align}
I(X;Y) = &p(X=1,Y=0)\log_{2}\frac{p(X=1,Y=0)}{p(X=1)p(Y=0)}+p(X=0,Y=1)\log_{2}\frac{p(X=0,Y=1)}{p(X=0)p(Y=1)} \nonumber \\
 & +p(X=1,Y=1)\log_{2}\frac{p(X=1,Y=1)}{p(X=1)p(Y=1)}+p(X=0,Y=0)\log_{2}\frac{p(X=0,Y=0)}{p(X=0)p(Y=0)} \nonumber \\
= & d\log_{2}\frac{d}{l(1-l)}+d\log_{2}\frac{d}{l(1-l)}+(l-d)\log_{2}\frac{l-d}{l^2}+(1-l-d)\log_{2}\frac{1-l-d}{(1-l)^2} \nonumber \\
= & 2d\log_{2}d-2l\log_{2}l-2(1-l)\log_{2}(1-l)+(l-d)\log_{2}(l-d)+(1-l-d)\log_{2}(1-l-d) \label{eq:xsmall}
\end{align}

To summarize,
\begin{equation*}
I(X;Y)=\left\{
	\begin{array}{l l}
		2(l-1)\log(1-l)+(1-2l)\log(1-2l) & \quad d\geq l \\
		2d\log_{2}d-2l\log_{2}l-2(1-l)\log_{2}(1-l)+(l-d)\log(l-d)+(1-l-d)\log(1-l-d) & \quad d<l\\
	\end{array}
\right.
\end{equation*}

\begin{figure}[!ht]
	\begin{center}
	\includegraphics[clip,width=10.0cm]{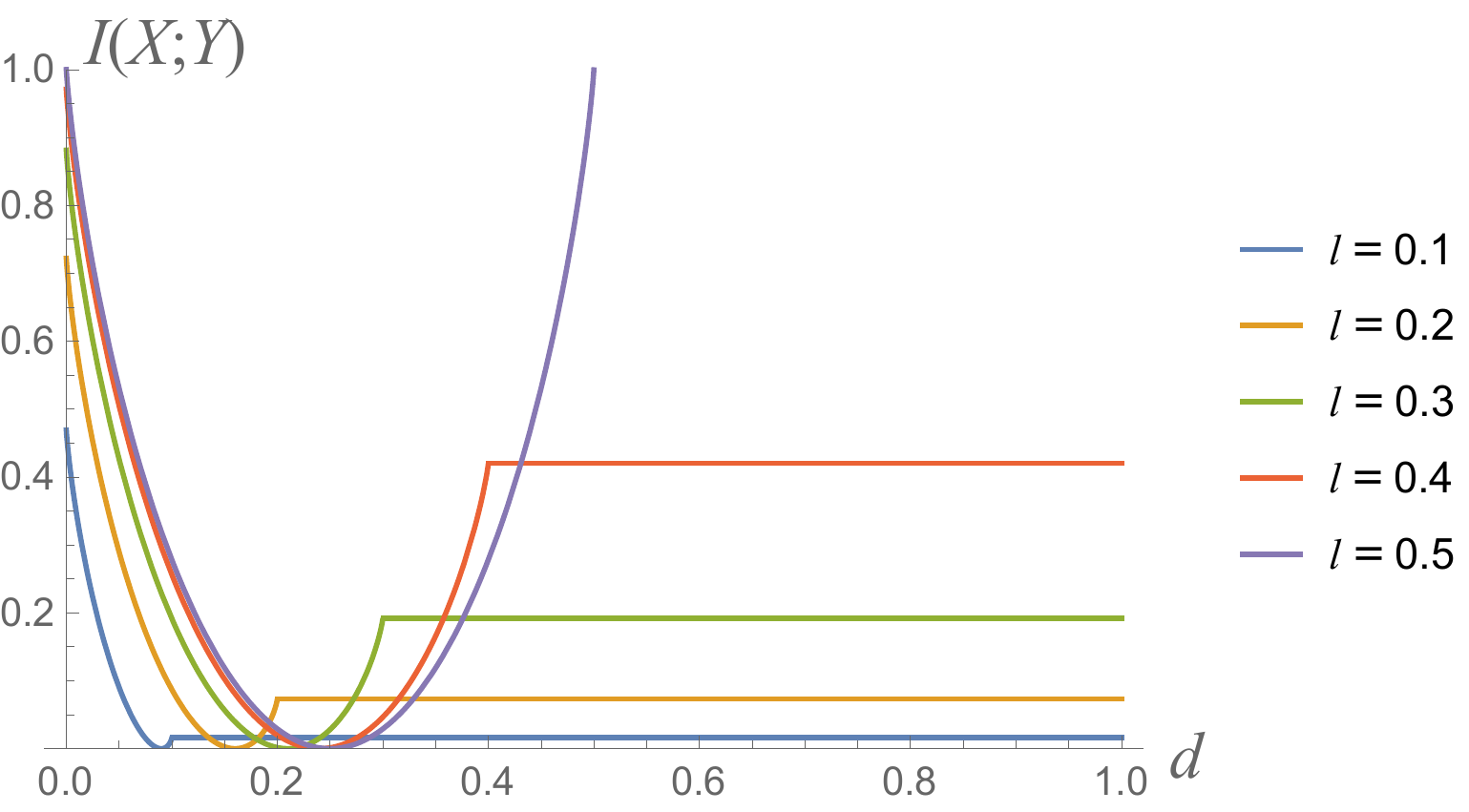}
	\end{center}
\caption{\footnotesize{\bf{Mutual Information Between Two Arbitrary Points as a Function of $d$ and $l$.}}}
\label{fig:mutual_info}
\end{figure}

From Equation \ref{eq:xlarge} and \ref{eq:xsmall}, as $DI$ increases, both $l$ and $d$ decrease. Accordingly, the mutual information between two arbitrary points at a distance $d$ becomes smaller (Figure \ref{fig:mutual_info}). This sensitivity of mutual information to $DI$ accounts for the lower average information sharing in the normal heartbeats in the cellular automata model compared to the reaction diffusion model in Figure 6 in the main paper. Figure \ref{fig:mutual_info} also illustrates the phenomenon of little information sharing when the two points are out of phase that is shown in the normal heartbeats in Figure 2 in the main paper.

\bibliography{references}